\documentclass[preprint2]{aastex}
\usepackage{amsmath}
\begin{document}
\title{Atmospheric scintillation at Dome~C, Antarctica: implications for photometry and astrometry}
\author{S.L. Kenyon, J.S. Lawrence, M.C.B. Ashley, J.W.V. Storey}
\affil{School of Physics, University of New South Wales, Sydney 2052, Australia}
\email{suzanne@phys.unsw.edu.au}
\author{A. Tokovinin}
\affil{Cerro-Tololo Inter American Observatory, Casilla 603, La Serena, Chile}
\and
\author{E. Fossat}
\affil{Laboratoire Universitaire d'Astrophysique de Nice, Universit\'{e} de Nice, France}
\shorttitle{Atmospheric scintillation at Dome~C, Antarctica}
\shortauthors{Kenyon, S.L. et al.}
\affil{}
\affil{Submitted to PASP 6th April 2006, accepted 26th April 2006}
\begin{abstract}
We present low-resolution turbulence profiles of the atmosphere above
Dome~C, Antarctica, measured with the MASS instrument during 25 nights
in March~--~May 2004. Except for the lowest layer, Dome~C has
significantly less turbulence than Cerro Tololo and Cerro Pach\'on. In
particular, the integrated turbulence at 16 km is {\it always less}
than the median values at the two Chilean sites. From these profiles
we evaluate the photometric noise produced by scintillation, and the
atmospheric contribution to the error budget in narrow-angle
differential astrometry. In comparison with the two mid-latitude sites
in Chile, Dome~C offers a potential gain of
about $3.6$ in both photometric precision (for long integrations) and
narrow-angle astrometry precision.  These  gain estimates
are preliminary, being computed with 
average  wind-speed profiles, but the validity of our
approach is confirmed by independent data. Although 
the data from Dome~C cover a fairly limited time frame, they lend
strong support to expectations that Dome~C will offer significant
advantages for photometric and astrometric studies. 
\end{abstract}

\section{Introduction}
The potential of the Antarctic plateau for astronomy
has been recognized for many years. In the early 1990's
\citet{Gillingham1991,Gillingham1993b} suggested that the atmospheric 
turbulence above a thin boundary layer would generally be very
weak. This was confirmed by the first measurements at the
South Pole by \citet{Marks1999}. These showed that most turbulence is found
close to the surface; in this case confined to a 200~m thick layer,
with very weak turbulence at high altitudes. In comparison, at most
temperate sites the turbulence is strong in the tropopause and above,
caused by the interactions of the jet stream with temperature
gradients in the tropopause. A brief history of astrophysics in
Antarctica is presented by \citet{Indermuehle2005}.

Dome~C, Antarctica is potentially one of the best astronomical
sites in the world. As a local maximum in elevation on the plateau,
Dome~C enjoys very low surface wind speeds, on average
2.9~m~s$^{-1}$\citep{Aristidi2005}. 

The atmospheric turbulence at Dome C has now been studied with four
different techniques: acoustic radar, MASS (a scintillation profiling
technique; see later), DIMM, and microthermal sensors.  An acoustic
radar, or SODAR, emits sound pulses into the air and derives the
strength of the atmospheric turbulence as a function of height from
the intensity and delay time of the reflected sound.  A DIMM, or
Differential Image-Motion Monitor, observes the relative motion of two
images of the same star viewed through two sub-apertures of a small
telescope.  From this, the DIMM can derive the integrated atmospheric
seeing.  Microthermal sensors, carried aloft on a weather balloon,
make an in situ measurement of the temperature fluctuations of the air
as a function of height all the way to the top of the atmosphere. 

SODAR  measurements in the early months of 2003 by
\citet{Travouillon2005} showed that, as expected, the surface
turbulent layer at Dome~C was much thinner that at South
Pole. Combined MASS and SODAR measurements of 
the turbulence in winter 2004 gave an average seeing of 0.27$\arcsec$
above 30 m, with the seeing below 0.15$\arcsec$ for 25\% of the time
\citep{Lawrence2004a}. DIMM measurements in winter 2005
\citep{Agabi2006} confirmed these results, showing an average seeing
of 0.25$\arcsec$ above the ground layer \citep[Fig. 1e]{Agabi2006}, and
balloon microthermal measurements by the same authors implies a median
seeing of 0.36$\arcsec\pm0.19\arcsec$ at a height of 30~m. They also
showed the existence of an intense turbulent boundary layer, finding a
median seeing of 1.9$\arcsec$ from ground level, with 87\% of the
total atmospheric turbulence confined to the first 36~m of atmosphere. 

As expected, at Dome C the sky background in the infrared
\citep{Walden2005} is lower than at temperate sites because of
the extremely cold temperatures and lower precipitable water
vapor. For the same reason, the atmospheric transmission in the
sub-millimeter is higher than at temperate sites \citep{Calisse2004}. 
An assessment of the optical sky brightness has been recently
published by \citet{Kenyon2006}.

To date, the winter time scintillation at Dome~C has only
been estimated from atmospheric models (e.g.,
\citealt{Swain2003}). Scintillation is an important factor in
measurements requiring high precision photometry (e.g., extra-solar
planet detection) and astrometry, and of objects with very fast
intensity changes (e.g. astroseismology;
\citealt{Heasley1996,Fossat2005}).  
 
Here we evaluate the scintillation noise contribution to photometry and the
atmospheric noise contribution to narrow-angle astrometry, using a set of
low-resolution turbulence profiles measured at Dome~C in 2004. The
instrument and data are described in \S~\ref{sec:MASS}. In
\S~\ref{sec:theory} we outline the theory of atmospheric turbulence, 
scintillation and interferometry. In \S~\ref{sec:results} we
present our results and discussion.

\section{MASS Measurements}\label{sec:MASS}
The turbulence profile of the atmosphere at Dome~C
was monitored with 
a MASS (Multi-Aperture Scintillation Sensor) instrument during the first 2
months of the 2004 night time: 23 March 2004 to 16 May 2004. The
analysis of these data in terms of seeing has been reported
by \citet{Lawrence2004a}.

The MASS instrument and theory are described in detail in
\citet{Kornilov2003} and \citet{Tokovinin2003b}. In brief, starlight
is directed via a telescope onto four concentric annular mirrors that
split the entrance aperture into rings with projected outer diameters
of 19, 32, 56, and 80~mm. Each of the four beams is directed to a
miniature Hamamatsu photomultiplier which samples the stellar
intensity at a 1~kHz rate. Four normal and six differential scintillation indices are
calculated for each 1~second integration and further averaged during
1~minute. The set of 10 indices is fitted to a model of six fixed layers
at heights 0.5, 1, 2, 4, 8 and 16~km above the observatory. For each
layer $i$, the integrated turbulence $J_i$~(m$^{1/3}$) is calculated:
\begin{equation}\label{eq:J}
  J_i=\int_{\textrm{Layer }i} C_{n}^{2}(h)\textrm{d}h,
\end{equation}
where $C_{n}^{2}(h)$~(m$^{-2/3}$) is the refractive index structure constant
 and $h$ is the height above the site. The spectral
response of MASS is from 400 to 550~nm with a FWHM bandwidth of 100~nm.

The  profile  restoration from  scintillation  indices  is a  delicate
procedure, and errors may reach 10\% of the total turbulence integral.
The errors are larger for the lower layers, while the 2 highest layers
(8~km and 16~km) are measured  well. The second moment of the
turbulence profile used in this paper  is measured by  MASS with high
reliability, as well as lower  moments.  This has been demonstrated by
inter-comparing MASS and SCIDAR instruments \citep{Tokovinin2005}.

The Dome~C MASS \citep{Lawrence2004b}, operated in the AASTINO
(Automated Astrophysical Site Testing International Observatory;
\citealt{Lawrence2005}), uses a gimbal-mounted siderostat mirror
feeding a fixed 85~mm refracting telescope. 

MASS instruments also operate at a number of other
sites. To provide a comparison to the Dome~C results, we have also
included the publicly available data\footnote{Obtained from the
National Optical Astronomy Observatory web page ``Sites Data Access,''
at http://139.229.11.21/}
from the Cerro Tololo and Cerro Pach\'on observatories in
Chile. The profiles for Cerro Pach\'on have been discussed and modeled
by \citet{TT2006}. 
\section{Theory}\label{sec:theory}
\subsection{Turbulence}
Many astronomical measurements are limited by the Earth's
atmosphere. A wavefront located at height $h$ and horizontal position 
vector ${\bf x}$ in the atmosphere can be described by its complex amplitude
$\Psi_{h}({\bf x})$ \citep{Roddier1981},
\begin{equation}
\Psi_{h}({\bf x})= \textrm{e}^{\chi_h({\bf x}) +  i\psi_{h}({\bf x})},
\end{equation}
where $\chi_h({\bf x})$ is the logarithm of the amplitude
and $\psi_{h}({\bf x})$ is the phase of the wave. 

Atmospheric turbulence introduces pure phase distortions. As the
wavefront propagates through the atmosphere, amplitude modulations
appear as well. In the geometric optics approximation, phase
perturbations act as positive or negative lenses, changing the
wavefront curvature and producing intensity modulation at the
ground. Diffraction is also important and defines the size of the most
effective atmospheric ``lenses'' to be of the order of the Fresnel radius,  
\begin{equation}
r_{F}\approx(\lambda h)^{1/2},
\end{equation}
where $\lambda$ is the wavelength of light and $h$ is the
height of the turbulent layer above the observatory site. For example,
if the dominant turbulence layer is at 10~km then
at 500~nm, $r_F=7~$cm.

\citet{Dravins1997a,Dravins1997b,Dravins1998} present detailed
discussions of stellar scintillation, including statistical
distributions and temporal properties, dependence on wavelength and
effects for different telescope apertures.  
\subsection{Scintillation noise}\label{sec:ScintilationIndex}
The scintillation index $\sigma_{I}^{2}$ is used as a measure of
the amount of scintillation and is defined (for small intensity fluctuations)
as the variance of $\Delta I/\left\langle I\right\rangle $. In the
weak-scintillation regime, $\sigma^2_I \ll 1$,  
the  effects of all turbulence layers are
additive. In this case, the 
scintillation index is related to the refractive index structure constant
$C_{n}^{2}(h)$ by \citep{Krause1993,Roddier1981}
\begin{equation}
\sigma_{I}^{2} = \int_{0}^{\infty}C_{n}^{2}(h)W(h)\textrm{d}h ,
\end{equation}
where the weighting function $W(h)$ is given by
\begin{equation}
  \begin{split}
    W(h) =& 16\pi^{2}0.033\left(\frac{2\pi}{\lambda}\right)^{2}\\
    \times&\int_0^\infty \left|A(f)\right|^2f^{-8/3}\sin^{2}
    \left(\frac{\lambda hf^{2}}{4\pi}\right)\textrm{d}f.
    \end{split}
\end{equation}
Here, $h$ is the height above the observatory, $\lambda$ is the
wavelength, $f$ is the spatial frequency and
$\left|A(f)\right|^2$ is an aperture filtering function. This
expression is valid for monochromatic light and has to be modified for
wide-band radiation. 

For telescope apertures with diameter $D\ll r_{F}$, the monochromatic
scintillation index is \citep{Roddier1981} 
\begin{equation}
\sigma_{I}^{2}  =
19.2\lambda^{-7/6}(\cos\gamma)^{-11/6}\int_{0}^{\infty}h^{5/6}C_{n}^{2}(h)\textrm{d}h\label{eq:SIzero},
\end{equation}
where $\gamma$ is the zenith angle.

The scintillation index for a large circular aperture with diameter
$D\gg r_{F}$ is: 
\begin{equation}\label{eq:SID}
\sigma_{I}^{2} =  17 D^{-7/3}(\cos\gamma)^{-3} \int_{0}^{\infty}h^{2}C_{n}^{2}(h)\textrm{d}h.
\end{equation}
Large apertures effectively average small-scale intensity
fluctuations, so that only atmospheric lenses of the order of the aperture
diameter~$D$ contribute to the flux modulation. In this case,
geometric optics applies and the scintillation becomes independent of
both the wavelength and the spectral bandwidth. 

The above expressions are for very short time scale exposures. For
exposure times that are longer than the time taken for a scintillation
pattern to cross the telescope aperture (i.e $t>(\pi D)/V_\perp$,
where $D$ the telescope diameter and $V_\perp$ the speed of
the turbulence layer), the scintillation index can be calculated from
\citep{Dravins1998} 
\begin{equation}\label{eq:sigmalargeT}
\sigma_I^2(t)=\int_0^\infty P(\nu){\rm sinc}^2(\pi \nu t){\rm d}\nu,
\end{equation}
where $\nu$~(s$^{-1}$) is the temporal frequency and $P(\nu)$ is the
temporal power spectrum, given by \citet[and ref. therein]{Yura1983} as  
\begin{equation}
P(\nu)\approx  8.27k^{2/3}\int_0^\infty\frac{C_n^2(h)h^{4/3}}{V_\perp(h)}Q(h){\rm d}h
\end{equation}
at the zenith, where $k=2\pi/\lambda$ and
\begin{eqnarray}\label{eq:Q}
Q(h)&=&\int_0^\infty
\left|A(f_x,f_y)\right|^2\\
&\times&\left[x^2+\frac {\nu^2}{\nu^2_0(h)}\right]^{-11/6}
\sin^2\left[x^2+\frac {\nu^2}{\nu^2_0(h)}\right]{\rm d}x.\nonumber
\end{eqnarray}
Here, $f_x=\nu/V_\perp(h)$,
 $f_y=(2k/h)^{1/2}x$, $\nu_0(h)=(2k/h)^{1/2}V_\perp(h)$ and $A(f)=(2J_1[fD/2])/(fD/2)$ for a circular aperture with diameter $D$ and $f^2=f_x^2+f_y^2$.

For large $t$, \citet{Dravins1998} simplify Equation
\ref{eq:sigmalargeT} to
\begin{eqnarray}
\sigma_I^2(t)&=&\frac{P(0)}{2t}.
\end{eqnarray}

In the limit of large apertures $D \ll r_F$, we can replace the sine
in Equation \ref{eq:Q} by its argument. By setting $\nu=0$ and introducing a
new variable $y = f_yD$, we can show that 
\begin{equation}
\label{eq:P0}
P(0) \approx 21.3D^{-4/3}\int_0^\infty\frac{C_n^2(h)h^{2}}{V_\perp(h)}{\rm d}h\;
\end{equation}
and is independent of the wavelength. 

For a particular set of turbulence and wind profiles, the
scintillation noise $\sigma_I$ at zenith can be expressed as
\begin{equation}\label{eq:sigma}
  \sigma_I = \begin{cases}
    S_1,          & D\ll r_F\\
    S_2 D^{-7/6}, & D\gg r_F\\
    S_3 D^{-2/3}t^{-1/2}, & D\gg r_F,\,t\gg(\pi D)/V_\perp\end{cases}
\end{equation}
where
\begin{eqnarray}
  S_1 & = & \left[19.2\lambda^{-7/6}
    \int_{0}^{\infty}h^{5/6}C_{n}^{2}(h)\textrm{d}h\right]^{1/2},\label{eq:S1}\\
  S_2   & = & \left[17.3  
    \int_{0}^{\infty}h^{2}C_{n}^{2}(h)\textrm{d}h\right]^{1/2},\label{eq:S2}\\
  S_3 & = & \left[10.7
    \int_0^\infty \frac{C_n^2(h)h^{2}}{V_\perp(h)}{\rm d}h\right]^{1/2}.
  \label{eq:S3}
\end{eqnarray}
The scintillation error can be expressed in magnitudes as
$\sigma_I({\rm mag})=2.5\log(\sigma_I+1)$.

In all cases the scintillation noise is dominated by the
high-altitude turbulence, more so in the case of large apertures
because of the $h^2$ weighting. It is the large-aperture case that is
generally of more relevance to astronomical photometry. 
\subsection{Astrometric interferometry}\label{sec:astrometry}
The Antarctic plateau has been recognized as a potentially favorable site for
interferometry because the high-altitude turbulence is very weak
\citep{Lloyd2002}. In particular, high precision, very-narrow angle differential
astrometry should be attainable at Dome~C using long baseline interferometry
techniques. This would benefit a number of science programs, including
extra-solar planet searches and the study of close binary and multiple
star systems (for other examples, see \citealt{Swain2003,Lloyd2002} and
\citealt{Sozzetti2005}).

Differential astrometric measurement requires simultaneous observations of the target and
reference object. To achieve this, each telescope has a dual feed to direct the beam
from each star to the beam combiner \citep{Shao1992}. On 
combination of the beams, a fringe pattern is produced if the
difference between the optical path lengths from each arm of the interferometer to the
beam combiner is within $\lambda^2/\Delta \lambda$ \citep{Lane2004}. 
The difference between the fringe positions of the two stars is
measured. Phase referencing can be used to improve the limiting
magnitude of the interferometer if the target star and reference
object are within the isoplanatic patch \citep{Shao1992}.

Uncertainties in astrometric position measurements arise from
instrumental effects (noise, systematic) and atmospheric effects
associated with temporal incoherence and anisoplanatism. See
\citet{Shao1992,Sozzetti2005} and \citet{Lane2004} for further
details.

The variance in an astrometric position measurement caused by
anisoplanatism (assuming a Kolmogorov turbulence spectrum) 
is described by \citet{Shao1992} as
\begin{equation}
  \sigma^2_{\rm atm} \approx 5.25t^{-1}\begin{cases}
    \theta^2 B^{-4/3}\int_{0}^{\infty}\frac{C_{n}^{2}(h)h^{2}}{V(h)}\textrm{d}h&\text{Case 1}\\
    \theta^{2/3}\int_{0}^{\infty}\frac{C_{n}^{2}(h)h^{2/3}}{V(h)}\textrm{d}h&\text{Case 2},\end{cases}
\end{equation}
where $t$ is the integration time, $\theta$ is the
angular separation between two stars, $C_{n}^{2}(h)$ and
$V(h)$ are the vertical turbulence and wind profiles,
$h$ is the height above the site and $B$ is the baseline or
diameter of the entrance pupil. These formulae are only approximate,
but the exact coefficient is not needed for the purpose of site
inter-comparison. 

Case 1 applies to interferometry when the integration time  $t\gg
B/\overline{V}$ and $\theta \overline{h}\ll B$; where $\overline{h}$
and $\overline{V}$ are the turbulence-weighted effective atmospheric
height and wind speed. Because of the $h^2$ weighting,
$\sigma_{\text{atm}}$ in this regime is very sensitive to
high-altitude turbulence. 

Case 2 is applicable to single dish astrometry and is independent of
the size of the telescope when $\theta \overline{h}\gg B$ and  $t\gg\theta
\overline{h}/\overline{V}$. 

For a particular set of turbulence and wind profiles, the error
$\sigma_\text{atm}$ (arcseconds) can be expressed as
\begin{equation}\label{eq:sigmaatm}
  \sigma_{\rm atm}=\begin{cases}
  C_1 t^{-1/2} \theta  B^{-2/3}&\text{Case 1}\\
  C_2 t^{-1/2} \theta^{1/3}&\text{Case 2} ,\end{cases}
\end{equation}
where
\begin{eqnarray}\label{eq:C1}
  C_1 & = & 472\;000\left[\int_{0}^{\infty}\frac{C_{n}^{2}(h)h^{2}}{V(h)}\textrm{d}h\right]^{1/2}
\end{eqnarray}
and
\begin{eqnarray}\label{eq:C2}
  C_2&=&472\;000\left[\int_{0}^{\infty}\frac{C_{n}^{2}(h)h^{2/3}}{V(h)}\textrm{d}h\right]^{1/2}.
\end{eqnarray}
Note that the expression for $C_1$ contains the same combination
of atmospheric parameters as the expression for the photometric
error $S_3$. This is not a coincidence, as both narrow-angle
astrometry and large-aperture photometry are affected by the same
physical phenomenon -- large-scale curvature fluctuations of
wavefronts. Hence, scintillation in large apertures contains information 
on the potential accuracy of narrow-angle astrometry at a given site. 

\section{Results}\label{sec:results}
Using the eight weeks of MASS data from Dome~C, we extracted 11532
turbulence profiles spread over 51 nights (we use ``night'' to mean
the period within 24 hours when the Sun is further than $10^\circ$
below the horizon). These data were filtered
according to the criteria: $B_D/F_D <0.03$, $F_D>F_\text{limit}$,
$\delta F_D < 0.003$, $0.7<F_C/F_D<0.9$ and $\chi^2<100$, where $F_D$
and $B_D$ are the star flux and background measurements in aperture D
(largest aperture). $F_\text{limit}$ is a cut-off
flux limit set to 100 counts for Alpha Trianguli, and 200 counts for
Beta Crucis and Beta Carinae. The parameter $\delta F_D$ shows slow flux
variations, used here to eliminate data affected by the guiding
errors. The flux ratio $F_C/F_D$ serves to control the aperture
vignetting by the entrance window, which was sometimes covered by snow
or frost. The $\chi^2$ is a measure of the fit quality. After
filtering, 1853 profiles over 26 nights remained for further analysis. 

Each profile includes the integrated turbulence   
$J_i$ (see Equation \ref{eq:J}) in layers centered at elevations of
0.5, 1, 2, 4, 8 and 16~km above the site with vertical resolution
$\Delta h/h\sim 0.5$. Finally, we calculated the scintillation noise
and implied astrometric error from each profile. In this section, we
compare these results with similar data for the Cerro Tololo and Cerro
Pach\'on observatories in Chile (see Table \ref{Tab:DataSets} for
information on each data set). 
\begin{deluxetable}{lcccccc}
\tablewidth{0pt}
\tablecolumns{7}
\tablecaption{Data sets\label{Tab:DataSets}}
\tablehead{    & &&&&\multicolumn{2}{c}{Number of }\\
\colhead{Site} & \multicolumn{2}{c}{Location} &\colhead{Altitude} &
  \colhead{Date range} & \colhead{Nights} &\colhead{Profiles} }
\startdata
Dome~C & $123^{\circ}21'$~E            & $75^{\circ}6'$~S  & 3260~m &  $23$   Mar 2005 -- $16$    May 2004  & $\phn26$  & $\phn1853$  \\
Cerro Tololo & $\phn70^{\circ}48'$~W   & $30^{\circ}9'$~S  & 2215~m & $19$    Mar 2002 -- $\phn2$ Feb 2006  & $573$     & $       98887$  \\
Cerro Pach\'on & $\phn70^{\circ}44'$~W & $30^{\circ}14'$~S & 2738~m & $\phn9$ Jan 2003 -- $30$    Jan 2006  & $293$     & $       39819$  \\
\enddata
\tablecomments{We use ``night'' to mean the period within 24 hours when the Sun is further than $10^\circ$ below the horizon}
\end{deluxetable}

\subsection{Turbulence profiles}
\begin{figure}
 \plotone{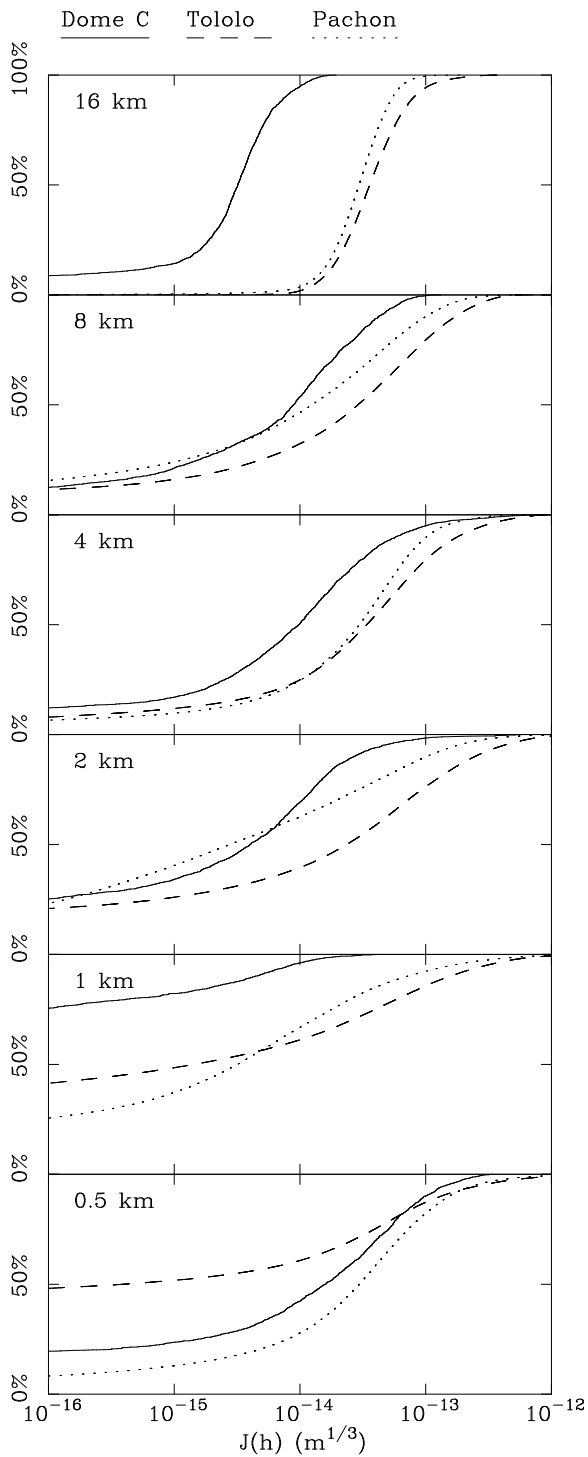}
\caption{\label{cap:Ji_hist} Cumulative
 probabilities that the integrated turbulence for each height (above
 the surface) is less
 than the given $J_i$ for Dome C ({\it solid}), Cerro Tololo ({\it dashed}) and Cerro Pach\'on ({\it dotted}). The large fraction of very low $J_i$ values for
 the 0.5~km and  1~km layers is an artifact of the MASS profile restoration method in
 situations when these layers do not dominate.}
\end{figure}

Figure \ref{cap:Ji_hist} shows the cumulative
probability that the integrated turbulence for each height is less
than the given $J_i$. 

Cerro Tololo has the lowest turbulence in the 0.5~km layer. At 1~km
the turbulence at Dome~C is so low that, for most of the time, it
cannot be reliably measured with MASS. Dome~C has a slightly higher
probability of smaller turbulence in the 4 and 8~km layers. However, the most
significant difference between the sites is in the 16~km layer; at
Dome~C the integrated turbulence in this high-altitude layer is
\emph{always less} than the median values at Cerro Tololo and Cerro Pach\'on.

The Cerro Tololo and Cerro Pach\'on sites are only 10~km apart and
have a 400~m altitude difference. Hence, we expect identical
high-altitude turbulence for those sites. The differences seen in
Figure  \ref{cap:Ji_hist} reflect  mostly  different seasonal coverage
of the data sets (more winter-spring data for Cerro Tololo) coupled to
the systematic seasonal trends in high-altitude turbulence. Similar
caution is warranted for the Dome~C data that cover only 25~nights.
\placefigure{cap:Ji_hist}
\subsection{Wind profiles}\label{sec:windprofiles}
The photometric error for long integration times (equations \ref{eq:sigma}
and \ref{eq:S3})
and the astrometric errors (equations \ref{eq:C1} and \ref{eq:C2}) depend
not only on the turbulence but also on the wind speed profile. As
these quantities are likely correlated, the correct way to 
estimate the errors requires simultaneous data on wind and
turbulence. The wind profiles can, in principle, be retrieved from the
global meteorological databases like NCEP (National Centers for
Environmental Prediction). However, here we adopt a
simplified approach and use fixed wind profile models instead. Hence,
the distributions derived here may be not realistic. 

Owing to the strong $h^2$ weighting, the astrometric and photometric
errors are almost entirely determined by the highest MASS layer at
16~km. Hence, the adopted wind speed in this layer critically
influences our results.  

The wind speed profiles for Cerro Pach\'on and Cerro Tololo were modeled using a
constant ground layer speed $V_g$ plus a Gaussian function to represent the jet
stream contribution \citep{Greenwood1977}.
\begin{equation}\label{eq:WindSpeed}
  V(h)=V_g + V_t \exp{\left[-\left(\frac{h-H}{T}\right)^2\right]} ,
\end{equation}
where $h$ is the altitude above the observatory. We set
$V_g=8$~m~s$^{-1}$, $V_t=30$~m~s$^{-1}$, $H=8$~km and $T=4$~km by
comparing the model to the Cerro Pach\'on wind profiles in \citet{Avila2000} and
\citet{Avila2001}. 

\begin{figure}[h]
  \plotone{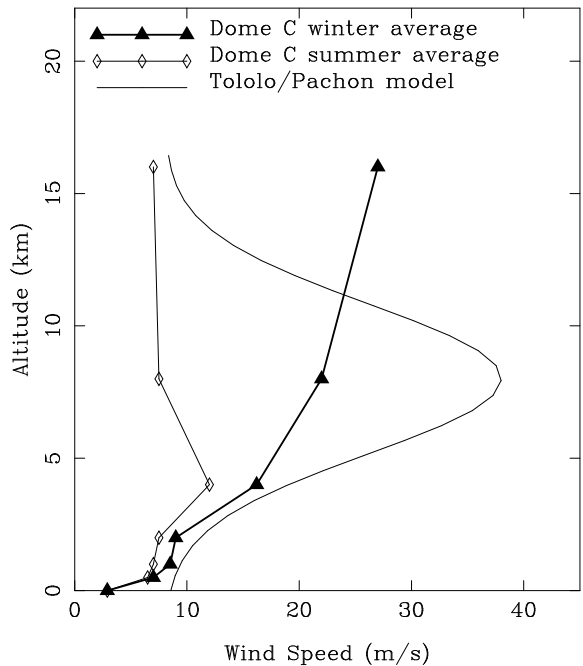}
  \caption{Average wind speeds profiles at Dome~C in the winter
    \citep{Agabi2006} and summer \citep{Aristidi2005}, and a model
    of the wind profile at Cerro Tololo and Cerro Pach\'on.\label{cap:DCWindSpeeds}}  
\end{figure}

The summer wind speed profile at Dome~C also shows a Gaussian peak at
the (somewhat lower) tropopause layer ($\sim 5$~km) and fairly
constant wind speed at other elevations (see Figure~4 of
\citealt{Aristidi2005}). In the winter, the wind speed profile is
different, showing an increase in stratospheric wind speeds and no peak at
the tropopause. So far only 3 profiles of the winter wind speed have
been published \citep{Agabi2006}. Figure~\ref{cap:DCWindSpeeds} shows
the average winter and summer time wind profiles, we used the
winter-time wind profile in this work.

\placefigure{cap:DCWindSpeeds}
\subsection{Scintillation noise}
\begin{figure}[p]
\plotone{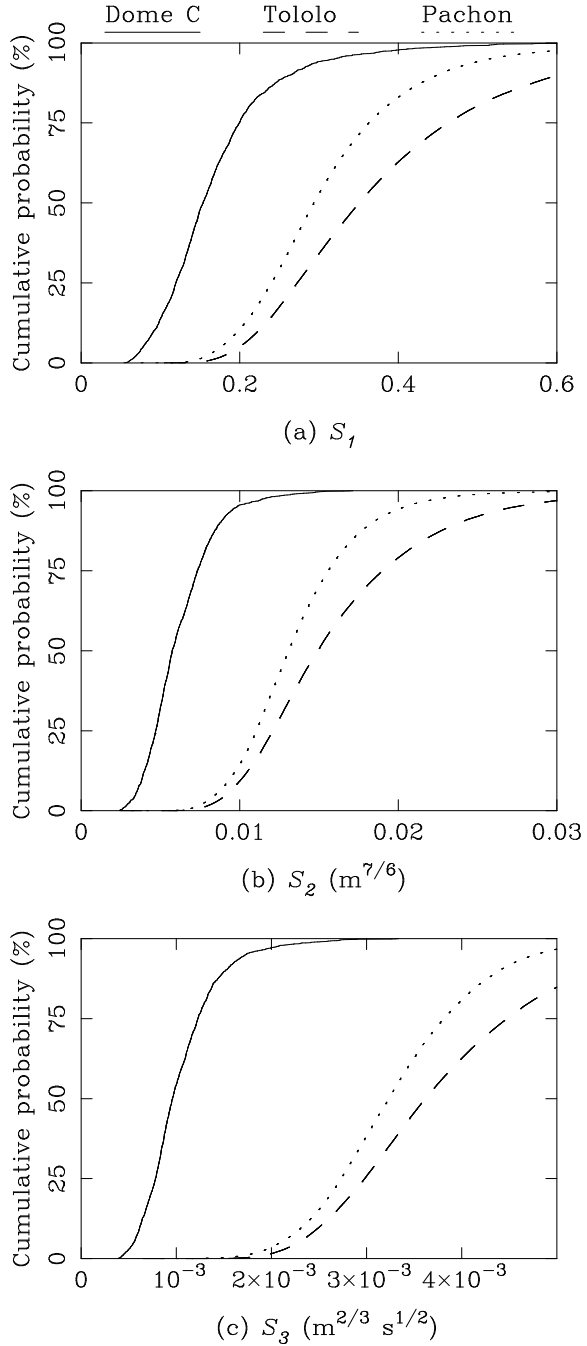}
\caption{\label{fig:SI}Cumulative probabilities of the constants
  (a) $S_1$, (b) $S_2$ (m${^7/6}$) and (c) $S_3$ (m$^{2/3}$~s$^{1/2}$) for Dome C, Cerro Tololo and
  Cerro Pach\'on. The scintillation noise $\sigma_I$ is the standard
  deviation of $\Delta I/I$ where $I$ is the stellar flux, and is equal
  to: $S_1$ for $D\ll 
  r_{F}$; $S_2D^{-7/6}$ for $D\gg r_F$, and $S_3D^{-2/3}t^{-1/2}$ for
  $D\gg r_F$ and $t\gg(\pi D)/V_\perp$, where $D$ is the telescope
  diameter, $r_F$ is radius of the Fresnal zone and $t$ is the integration time.}
\end{figure}

The scintillation noise was calculated for the three regimes discussed
in \S~\ref{sec:ScintilationIndex}, using the results from
each site. Figure~\ref{fig:SI} shows the cumulative probabilities for
$S_1$, $S_2$ and $S_3$. 

For short time scales, the median scintillation noise at Dome C
is a factor of $\sim 2$ less than at Cerro Tololo and Cerro Pach\'on, in the small
aperture regime. For larger apertures the gain is slightly higher,
$\sim 2.4$, because of the weaker high altitude turbulence at
Dome~C. As an example, for a 4~m diameter telescope, the median values
of the scintillation noise at each site are: 1.2~mmag (Dome~C),
3.2~mmag (Cerro Tololo) and 2.8~mmag (Cerro Pach\'on). 

The more relevant figure is the scintillation noise for
long exposure times. We used the wind speed models discussed above, to
calculate this parameter. Dome~C offers a potential gain of about
3.6 in photometric precision compared to Cerro Tololo and Cerro Pach\'on. From the
results we calculate the median photometric error expected on a 4~m
telescope for $t=60~$s to be: $\sim53~\mu$mag, at
Dome~C; $\sim200~\mu$mag at Cerro Tololo, and $\sim180~\mu$mag at Cerro Pach\'on.   

As a comparison, \citet{Dravins1998} measured $P(\nu)$ at La
Palma using various small apertures. From their results they
extrapolate $P(0)=5\times10^{-6}$~s for a 4~m aperture at zenith,
which gives $\sigma_I=220~\mu$mag for a 60~s integration, similar to the
typical values for Cerro Tololo and Cerro Pach\'on. Our results are also consistent
with those measured at Kitt Peak and Mauna Kea by
\citet{Gilliland1993}. 

\subsection{Astrometry}

\begin{figure}
\plotone{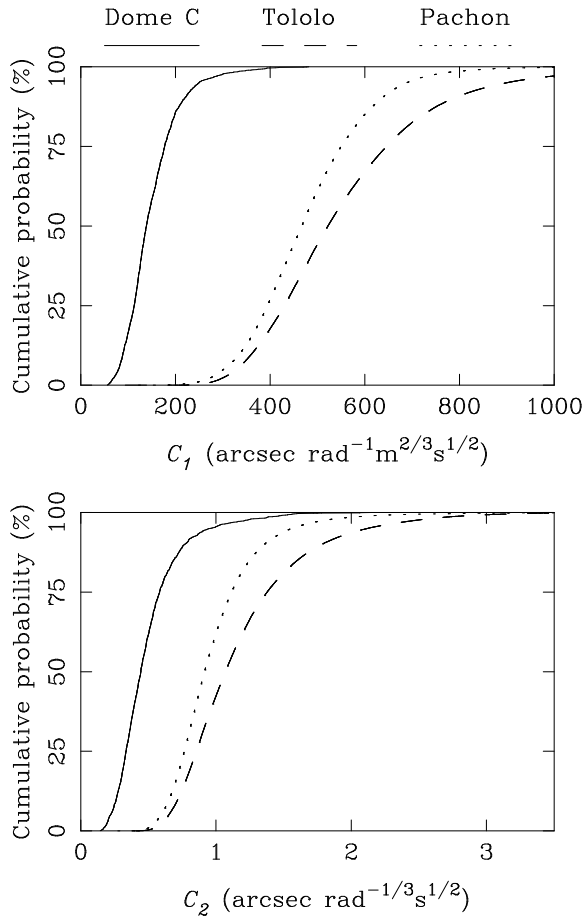}
\caption{\label{fig:HistC}Cumulative probabilities that the constants
  $C_1$ and $C_2$ are less than the value given, for Dome~C, Cerro
  Tololo and Cerro Pach\'on. The astrometric error $\sigma_{\rm atm}$ (arcsecond)
  is equal to: $C_1t^{-1/2}\theta B^{-2/3}$ for $t\gg B/\overline{V}$
  and $\theta\overline{h}\ll B$, and $C_2t^{-1/2}\theta^{1/3}$ for
  $t\gg \theta\overline{h}/\overline{V}$ and $\theta\overline{h}\gg
  B$. Here, $t$ is the integration time, $B$ is the baseline length, $\overline{h}$
and $\overline{V}$ are the turbulence-weighted effective atmospheric
height and wind speed, and $\theta$ is the stellar separation. }
\end{figure}

The constants $C_{1}$ and $C_{2}$ (\S~\ref{sec:astrometry}) were
calculated for each site; cumulative probabilities are shown in
Figure~\ref{fig:HistC}. The median astrometric error
$\sigma_{\text{atm}}$ at 
Dome~C is $\sim 3.5$ times less than the median values at Cerro Tololo
and Cerro Pach\'on. In Figure~\ref{fig:sigma_{delta}},
$\sigma_{\text{atm}}$ at Dome~C is plotted for several baselines, as a  
function of separation angle~$\theta$ for an integration time of 1~h.

We note that the  advantage of Dome~C for narrow-angle astrometry
over mid-latitude sites is even  larger than its advantage in the fast
scintillation.  This difference  is related  to the  adopted
wind  speed  at 16~km  altitude (27~m~s$^{-1}$  and 8.5~m~s$^{-1}$
for  Dome~C and  Cerro Pach\'on, respectively). Turbulence at Dome~C
is known to be slow (large time constant, see below), but the $h^2$
weighting in the expressions for the photometric and astrometric
errors reverses this conclusion because the high altitude turbulence
dominates the calculation. 

Our conclusions are conditional on the adopted  wind-speed models.
Using the mean Dome~C winter wind speed we calculated a median
$C_1$ value of 140~arcsec~rad$^{-1}$m$^{2/3}$s$^{1/2}$; decreasing the
16~km wind speed to 7~m~s$^{-1}$ gave a median $C_1$ value of
200~arcsec~rad$^{-1}$m$^{2/3}$s$^{1/2}$, still well below the median
values at Cerro Tololo and Cerro Pach\'on. As an additional check, we
computed $C_1$ from a set of six balloon profiles of $C_n^2$ and
wind   measured  at Cerro Pach\'on   in   October  1998
(see \citealt{Avila2000} and \citealt{Avila2001} for  the discussion
of these data). The $C_1$ values range from 380 to
660~arcsec~rad$^{-1}$m$^2$s$^{1/2}$, with a median of 480. This is
close to the median value for Cerro Pach\'on given in
Figure~4. \citet{Shao1992} calculate $C_1$ at Mauna Kea to be
300~arcsec~rad$^{-1}$m$^2$s$^{1/2}$, using the results from 2 short
observing campaigns.

The fringe phase of an interferometric measurement must be determined
within the atmospheric coherence time. Table \ref{tab:siteparameters}
shows the median coherence times $\tau_0$ \citep{Roddier1982a} and
isoplanatic angles $\theta_0$ \citep{Roddier1982b} at each site for
wavelengths 500~nm and 2.2~$\mu$m. The median coherence time at
Dome~C, measured with MASS, is 7.2~ms at $\lambda=500$~nm and 42~ms at
$\lambda=2.2~\mu$m. Using the measured turbulence profiles and assumed wind
profile at Dome C we calculated $\tau_0=9.4$~ms at $\lambda=500$~nm.

Phase referencing during the measurement
\citep{Shao1992} effectively increases the coherence time, with the
condition that the target and reference objects are within the same
isoplanatic patch. The median isoplanatic angle at
Dome~C is $\sim 3$ times larger than at Cerro Tololo and Cerro Pach\'on, allowing
wider fields to be used for phase referencing.
\begin{figure}
\plotone{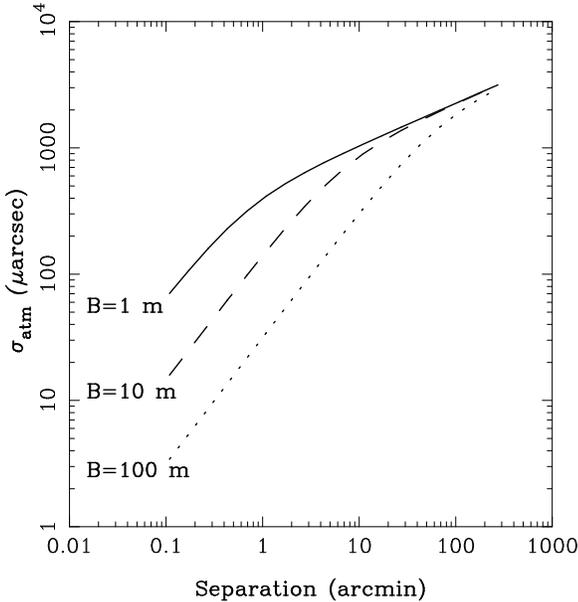}
\caption{\label{fig:sigma_{delta}}The median error $\sigma_{\text{atm}}$ for
  three baseline lengths at Dome~C, with an integration time of 1~h.}
\end{figure}
\begin{deluxetable}{lcccc}
  \tablewidth{0 pt}
  \tablecolumns{5}
  \tablecaption{\label{tab:siteparameters}Isoplanatic angles and
  coherence times for adaptive optics}
  \tablehead{\colhead{Site}
    &\colhead{$\theta_0$(500 nm)} 
    &\colhead{$\tau_0$(500 nm)}
    &\colhead{$\theta_0$(2.2 $\mu$m)} 
    &\colhead{$\tau_0$(2.2 $\mu$m)}\\
    \colhead{} 
    &\colhead{arcsec} 
    &\colhead{ms}
    &\colhead{arcsec} 
    &\colhead{ms}}
  \startdata
    \multicolumn{5}{c}{\emph{Median values}}\\
  Dome~C &  5.4 &  7.2 &  32 & 42 \\
  Cerro Tololo &  1.8 &  2.0 &  11 & 12 \\
  Cerro Pach\'on &  2.0 &  2.6 &  12 & 16 \\
    \multicolumn{5}{c}{\emph{Average values}}\\
   Dome~C &  5.9 &  8.8 &  35 & 52 \\
  Cerro Tololo &   1.9 &  2.8 &  11 & 17 \\
  Cerro Pach\'on &  2.1 &  3.3 &  13 & 20 
  \enddata
\tablecomments{$\tau_0$ and $\theta_0$ at 500~nm are taken from the MASS data
  files, the values at 2.2~$\mu$m are scaled by $\lambda^{6/5}$.} 
\end{deluxetable}
\placefigure{fig:sigma_{delta}}
\placetable{tab:siteparameters}

\section{Conclusions}
The scintillation noise at Dome~C for fast exposures is typically a
factor $1.9$~--~$2.6$ times lower than at Cerro Tololo and Cerro
Pach\'on; leading to a corresponding reduction in this ultimate limit
for high precision photometry. The ``small aperture'' scintillation
index becomes important for adaptive optics when the distance between
the actuators approaches the Fresnel zone size $r_F$, and shadow patterns start
becoming resolved \citep{Masciadri2004}. Adaptive optics will also
benefit from the long coherence time and large isoplanatic angle at
Dome~C, particularly in the infrared. 

For longer exposures, $\sigma_I$ at Dome~C is
typically $3.4$~--~$3.8$ times less than at Cerro Tololo and Cerro Pach\'on. For a
60~second integration on a 4~m telescope, the median  
photometric error is $\sim 53~\mu$mag at Dome~C. This parameter is
important for exoplanet transit measurements because the change in flux,
caused by a transiting planet, is related to the planet $R_p$ and star
$R_s$ radii by $\Delta F/F=(R_P/R_S)^2$. For example, for a Jupiter size
planet transiting a Sun size star $\Delta F/F=0.01$, for an Earth size
planet this ratio is 0.0001. The lower scintillation noise at Dome C
will allow for the transits of smaller planets to be detected than at
the Chilean sites. 

The atmospheric contribution to the positional error in a differential
astrometric measurement using a long baseline interferometer at Dome~C
is always less than the median values at Cerro Tololo and Cerro Pach\'on. This
conclusion is obtained using average wind profiles and remains
provisional until a more complete analysis is done. 

Based on the expected low astrometric error at 
Dome~C, a number of interferometric projects have already been proposed.
These include the Antarctic Planet Interferometer (API;
\citealt{Swain2004}), the Kiloparsec Explorer for Optical Planet
Search (KEOPS; \citealt{Vakili2004}) and the Antarctic L-band
Astrophysics Discovery Demonstrator for Interferometric Nulling
(ALADDIN; \citealt{Coude}). Many science programs would
benefit from an Antarctic interferometer, including exoplanet detection and orbit
determination and measurement of micro-lensing events \citep{Lloyd2002}.  

\acknowledgments
The UNSW AASTINO project is indebted to the French and Italian
Antarctic Programs (IPEV, PNRA) for logistics support and to the
Australian Research Council and the Australian Antarctic Division for
financial support. SLK is supported by an
Australian Postgraduate Award and by an Australian Antarctic Division
top-up scholarship. We thank James Lloyd and Mark Swain for
particularly useful  discussions that led to the concept of a ``warm
MASS'' for making these  measurements, and Victor Kornilov and Nicolai
Shatsky for helpful advice on the data reduction  software. We thank
the LUAN team at the University of Nice; in particular Eric Aristidi and
Karim Agabi for their assistance in setting up the AASTINO at
Dome~C. We also thank Colin Bonner, Jon Everett and 
Tony Travouillon of the University of New South Wales Antarctic
Research Group, and Anna Moore of the Anglo-Australian Observatory, for
valuable contributions to the Dome~C MASS project. 

\bibliographystyle{astron}


\newpage

\end{document}